\documentclass[conference, twocolumn, 10pt]{IEEEtran}

\usepackage{graphicx}
\usepackage{cite}
\usepackage{amssymb,amsmath}
\usepackage{amsthm}
\usepackage{algorithm}
\usepackage{algpseudocode}
\usepackage{color}
\usepackage{epstopdf}

\usepackage{subcaption}

\newtheorem{lemma}{Lemma}

\newtheorem{definition}{Definition}

\begin{document}
\title{\huge{ Improper Signaling for Virtual Full-Duplex Relay Systems }}
\author{{ Mohamed Gaafar\IEEEauthorrefmark{2}, Osama Amin\IEEEauthorrefmark{3}, Rafael F. Schaefer\IEEEauthorrefmark{2}, and Mohamed-Slim Alouini\IEEEauthorrefmark{3}} 
\\[0.25cm]
\IEEEauthorrefmark{2} Information Theory and Applications Chair, Technische Universit\"at Berlin, Germany\\
\IEEEauthorblockA{\IEEEauthorrefmark{3}Computer, Electrical, and Mathematical Sciences and Engineering (CEMSE) Division\\
King Abdullah University of Science and Technology (KAUST), Thuwal, Makkah Province, Saudi Arabia.
\\[0.25cm]
E-mail: {\{{mohamed.gaafar, rafael.schaefer\}@tu-berlin.de}}, {\{{osama.amin, slim.alouini\}@kaust.edu.sa}}
}}
\maketitle
\pagenumbering{gobble}
\begin{abstract}
Virtual full-duplex (VFD) is a powerful solution to compensate the rate loss of half-duplex relaying without the need to full-duplex capable nodes. Inter-relay interference (IRI) challenges the operation of VFD relaying systems. Recently, improper signaling is employed at both relays of the VFD to mitigate the IRI by imposing the same signal characteristics for both relays. To further boost the achievable rate performance, asymmetric time sharing VFD relaying system is adopted with different improper  signals at the half-duplex relays. The joint tuning of the three design parameters improves the achievable rate performance at different ranges of IRI and different relays locations. Extensive simulation results are presented and analyzed to show the achievable rate gain of the proposed system and understand the system behavior. 

\end{abstract}

\section{Introduction}

Cooperative relaying has been implmented in several standards to improve the communication links performance and reliablity in wireless networks \cite{Yang_HXM_COMMAG09}. Different techniques  are adopted to develop the cooperative relaying operation such as relay selection and resource allocation \cite{amin2012adaptive, amin2011optimal}. The current implementations consider half-duplex (HD) relays that transmit and receive in different time or frequency slots. To compensate the spectral efficiency loss in HD relaying cooperative systems, full-duplex (FD) relaying systems are used double the spectral efficiency by allowing the relays to transmit and receive simultaneously in the same frequency band. However, implementing  
FD relaying systems requires the availability of new transceiver nodes with special hardware requirements  to overcomes the strong self-interference \cite{sabharwal2014band}.


Virtual full-duplex (VFD) relaying system was proposed in \cite{Rankov_W_JSAC07} to realize the FD  relaying system in a distributed manner without upgrading the network hardware\footnote{VFD is known also as \emph{two-path relaying} or \emph{alternate relaying}.}. VFD uses two HD relays to transmit and receive in turn, i.e.,  the source (destination) transmits (receives) a message in every time slot similar to the FD mode 
while being subjected to less interference than FD relaying. 
The interference source in VFD is the HD relay that transmits its signal to the destination, meanwhile the other relay receives this interfered signal which is known in literature by Inter-relay interference (IRI). The operation of VFD relaying scheme can be limited by the IRI  unless suitable interference mitigation techniques are used  \cite{ju2009catching}. 
 
Improper  signaling  is recetly proved its effectivness to reduce the interference impact on different communication systems \cite{gaafar2017improper, Lameiro_SS_WCL15, amin2017overlay, gaafar2017underlay}. Improper signaling is a asymmetric Gaussian signaling scheme with unequal power of the real and imaginary components and/or dependent real and imaginary components. As such, it has the ability to control the interference signature in the signal dimension, and thus it is considered as kind of signal interference alignment \cite{kurniawan2015improper}. Recently, we considered  the VFD relaying problem and we showed that the IRI can be significantly alleviated if the HD relays transmit improper signals \cite{gaafar2016letter, gaafar2017improper}.  Specifically, in \cite{gaafar2016letter}, we assumed that both relays use the improper signals with the same statistical characteristics and adjust the degree of impropriety which is measured by the so-called circularity coefficient. On the other hand, in \cite{gaafar2017improper}, we optimized both the power and the  circularity coefficient of the improper signal assuming using the same signal characteristics of both relays. Although imposing the same signal characteristics of both relays resulted in a notable gain for VFD in \cite{gaafar2016letter, gaafar2017improper} compared with the the traditional symmetric Gaussian signaling, optimizing the statistical characteristics for each relay would achieve more gain than fixed relays characteristics especially for asymmetric relay locations.

In this paper, we introduce a generalize study of \cite{gaafar2016letter} that allows the relays to use different statistical characteristics in order to maximize the benefits of the improper Gaussian scheme. Moreover, to further boost the rate performance of the VFD, we assume asymmetric VFD system that uses unequal transmission phases. 

\section{Virtual Full-Duplex Relaying System with Asymmetric Time Sharing }\label{sec:sys_mod}

\begin{figure*}[!t]
\centering
\includegraphics[width= 5.5in]{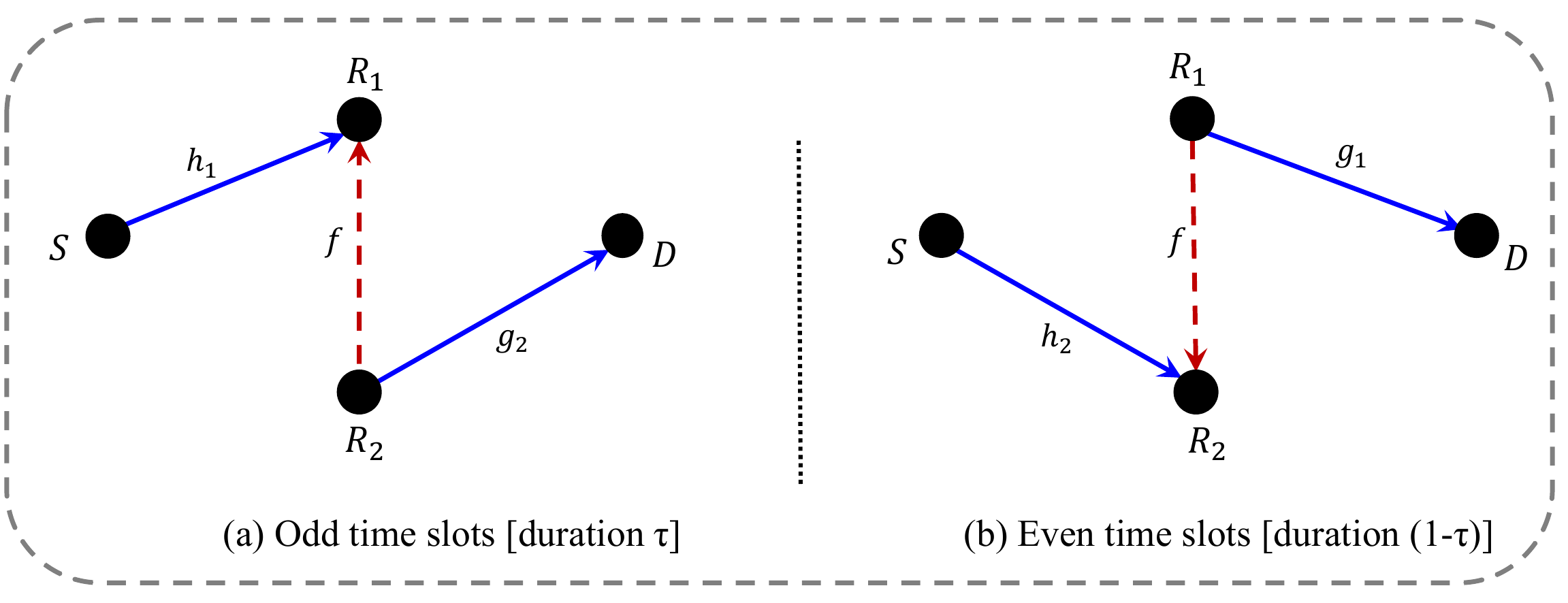}
\caption{VFD relay network. The blue solid lines represent the signal links and the red dashed lines represent the IRI links.}
\label{fig1}
\end{figure*}

Consider a VFD relaying network consisting of one source node, $S$, two HD relay nodes, $R_1$ and $R_2$, and one destination node, $D$, as shown in Fig. \ref{fig1}. The relays transmit and receive in turn, i.e., one relay relay transmits and the other one receives in one time slot, while the scenario is reversed in the next time slot. Moreover, we assume asymmetric time slots by considering the odd time slots with a normalized duration of $\tau$ and odd time slots with a normalized duration of $1-\tau$. For equal time sharing between the two transmission phases, we have $\tau=0.5$. As for the relaying protocol, we assume decode-and-forward (DF) at both relays. Let $h_i$ and $g_i$, $i \in \{1, 2\}$, denote the channel between $S$ and $R_i$, and the channel between $R_i$ and $D$, respectively. We assume channel reciprocity for the inter-relay channel, which is denoted by $f$. Moreover, we assume the source transmit power is $p_{\rm{s}}$, the relay transmit power is $p_{\rm{r}}$, and the noise variance at each receiving node is $\sigma_{{n}}^2$. The transmit power terms are limited to a power budget of $p_{\rm{max}}$.
During time $k$, the signal received at $R_i$ with $i=2-\mod(k,2)$ is given by\footnote{For the rest of the paper, we let $j,i \in \{1,2\}$, $j \neq i$}
\begin{equation}\label{eq1}
    y_i[k] = \sqrt{p_s} h_i[k] s[k]+\sqrt{p_r}f[k] x_{j}[k]+n_i[k],
\end{equation}
where $s[k]$ is the transmit proper signal by $S$  in time slot $k$ and $n_i[k]$ is the additive noise at $R_i$ with variance $\sigma_{\rm{n}}^2$. $x_j[k]$ is the improper signal, with circularity coefficient $\mathcal{C}_j$, transmitted by $R_j$ with $j=1+\mod(k,2)$. The received signal at $D$ from $R_i$ in time slot $k+1$ is given by
\begin{equation}\label{eq2}
    y_D[k+1] = \sqrt{p_r}g_i[k+1] x_{i}[k+1]+n[k+1],
\end{equation}
where $n[k+1]$ is the additive noise at $D$.
In the following, we assume that the channels are assumed to be quasi-static block flat fading channels, and therefore we drop the time index $k$ for notational convenience. The additive noise at the receivers is modeled as a white, zero-mean, circularly symmetric, complex Gaussian with variance $\sigma_{{n}}^2$.

Similar to \cite{gaafar2016letter}, we assume that no channel state information is available at $S$ and hence dirty paper coding cannot be used to fully cancel the IRI. Also, we assume a fixed transmit power at the relays and no direct link is available between $S$ and $D$. However, on the contrary to \cite{gaafar2016letter}, we assume here that the relays have different circularity coefficients and the relays' transmission phases could be unequal,  which are expected to give a better performance than in \cite{gaafar2016letter} where we assume the same circularity at both relays for tractability and simplification along with equal time slots duration. The reason of the expected advantage here is that our new model here provides more degrees of freedom in the design problem than can be exploited to mitigate the IRI. 

\section{Virtual Full Duplex Performance} 
In this section, we express the achievable rate performance of the VFD system when the relays employ different improper signals. Moreover, the VFD uses different duration of relaying phases. Before analyzing the achievable rate of the proposed system, we introduce the following basics about improper signaling. 

\subsection{Improper Signaling}
 First, we consider a zero mean random Gaussian variable $x$ definitions of improper Gaussian random variables. %

\begin{definition}  \label{def1}
The variance and the complementary (pseudo-) variance of  $x$ is defined, respectively, ass ${\sigma}_x^2=\mathbb{E} \{x^2\}$  and $\tilde{\sigma}_x^2=\mathbb{E} \{x^2\}$,   where $\mathbb{E}\{.\}$ denotes the expectation operator \cite{neeser1993proper}. 
\end{definition}

\begin{definition} \label{def2}
The Gaussian signal $x$ is called proper if $\tilde{\sigma}_x^2=0$, otherwise it is called improper \cite{neeser1993proper}.
\end{definition}
\begin{definition} \label{def3}
The circularity coefficient of the signal $x$ is a measure of its impropriety (asymmetry) degree and is defined as $\mathcal{C}_x ={|\tilde{\sigma}_x^2|}/{\sigma_x^2}$, where ${\sigma}_x^2=\mathbb{E}\{|x|^2\}$ is the conventional variance and $|.|$ is the absolute value operation \cite{Lameiro_SS_WCL15}. 
\end{definition}
From the aforementioned definitions, the circularity coefficient is bounded such that $0\leq \mathcal{C}_x \leq 1$.  In particular, $\mathcal{C}_x = 0$ and $\mathcal{C}_x = 1$ correspond to proper and maximally improper signals, respectively.

\begin{lemma} \label{lemma3}
The achievable rate expression of a single link that is subjected to interference and noise  $z$ when $x$ is transmitted and observed as $y$ is expressed as \cite{zeng2013transmit}
\begin{equation} 
R = \frac{1}{2}\log_2 \left( {\frac{{\sigma _{{{{y}}}}^4 - {{\left| {\tilde \sigma _{{{{y}}}}^2} \right|}^2}}}{{\sigma _z^4 - {{\left| {\tilde \sigma _z^2} \right|}^2}}}} \right).
\end{equation}
\end{lemma} 

\subsection{Achievable Rate Performance}
As a result of using improper signals at $R_j$ and proper signals at $S$ in addition to using unequal time slots durations, while treating the interference as Gaussian noise, the achievable rates of the VFD system is analyzed based on Lemma 1. First, the first hop of the $i$th path ($S-R_i$) can be expressed after some simplification steps as  \cite{zeng2013transmit,gaafar2016letter}
\begin{align}
{\mathcal{R}_{i,1}}\left( {{{\cal C}_j}} \right) &= \frac{1}{2} \times \nonumber \\
 &{\log _2}\left( {1 + \frac{{2{p_{\rm{s}}}{{\left| {{h_i}} \right|}^2}\left( {{p_{\rm{r}}}{{\left| f \right|}^2} + \sigma _n^2} \right) + p_{\rm{s}}^2{{\left| {{h_i}} \right|}^4}}}{{\left( {1 - {\cal C}_j^2} \right)p_{\rm{r}}^2{{\left| f \right|}^4} + 2{p_{\rm{r}}}{{\left| f \right|}^2}\sigma _n^2 + \sigma _n^4}}} \right).
\end{align} 
Similarly, the achievable rate of the second hop of the $i$th path can be obtained from \eqref{eq2} as
\begin{equation}
{\mathcal{R}_{i,2}}\left( {{{\cal C}_i}} \right) = \frac{1}{2}{\log _2}\left( {1 + \frac{{2{p_{\rm{r}}}{{\left| {{g_i}} \right|}^2}}}{{\sigma _n^2}} + \frac{{p_{\rm{r}}^2{{\left| {{g_i}} \right|}^4}\left( {1 - {\cal C}_i^2} \right)}}{{\sigma _n^4}}} \right).
\end{equation} 
Accordingly, the overall end-to-end achievable rate of the two-path relaying system is expressed as 
\begin{equation}\label{R_tot}
{\mathcal{R}_{{\rm{T}}}}\left( {{{\cal C}_1},{{\cal C}_2},\tau} \right) = \sum\limits_{i = 1}^2 {{R_{{_i}}}\left( {{{\cal C}_1},{{\cal C}_2},\tau} \right)},
\end{equation} 
where
\begin{equation}
{\mathcal{R}_{{_i}}}\left({{{\cal C}_1},{{\cal C}_2},\tau} \right) = \min \Big\{ {\tau{\mathcal{R}_{i,i}\left( {{{\cal C}_2}} \right)},{(1-\tau)\mathcal{R}_{i,j}\left( {{{\cal C}_1}} \right)}} \Big\}.
\end{equation}
For equal time sharing $\tau=0.5$, the effective achievable rate of the $i$th path, ${\mathcal{R}_{{_i}}}$ can be simplified after some algebraic steps giving \eqref{R_i_branches} that appears on the top of the following page.
\begin{figure*}[!t]
 \begin{align}\label{R_i_branches}
\begin{array}{l}
{{\cal R}_i}\left( {{{\cal C}_i},{{\cal C}_j},0.5} \right) = \min \left\{ {{{\cal R}_{i,1}}\left( {{{\cal C}_j}} \right),{{\cal R}_{i,2}}\left( {{{\cal C}_i}} \right)} \right\}\\
\\
\qquad \qquad \qquad \;\;  = \left\{ {\begin{array}{*{20}{l}}
{{{\cal R}_{i,1}}\left( {{{\cal C}_j}} \right),\quad {\rm{if}}}&\begin{array}{l}
{{\cal R}_{i,1}}\left( 1 \right) \le {{\cal R}_{i,2}}\left( 1 \right)\quad {\rm{or     }}\\
{{\cal C}_i} \le  {\sqrt {1 - \Psi \left( {1 - {\cal C}_j^2} \right)} } ,{{\cal R}_{i,1}}\left( 1 \right) > {{\cal R}_{i,2}}\left( 1 \right)\;,\;{{\cal R}_{i,2}}\left( 0 \right) > {{\cal R}_{i,1}}\left( 0 \right)
\end{array}\\
\\
{{{\cal R}_{i,2}}\left( {{{\cal C}_i}} \right),\quad {\rm{if}}}&\begin{array}{l}
{{\cal R}_{i,2}}\left( 0 \right) \le {{\cal R}_{i,1}}\left( 0 \right)\quad {\rm{or}}\\
{{\cal C}_i} \ge {\sqrt {1 - \Psi \left( {1 - {\cal C}_j^2} \right)} } ,{{\cal R}_{i,1}}\left( 1 \right) > {{\cal R}_{i,2}}\left( 1 \right)\;,{{\cal R}_{i,2}}\left( 0 \right) > {{\cal R}_{i,1}}\left( 0 \right)
\end{array}
\end{array}} \right.
\end{array}
 \end{align}
 \\\\
 where
 $\Psi \left( x \right) = \frac{{\sigma _n^2}}{{{p_{\rm{r}}}{{\left| {{g_i}} \right|}^2}}}\left( {\frac{{\gamma  - \alpha x}}{{\beta  + \alpha x}}} \right)$ and\\\\\\
$\alpha  = \frac{{2p_{\rm{r}}^3{{\left| {{g_i}} \right|}^2}{{\left| f \right|}^4}}}{{\sigma _n^2}},\;\beta  = {p_{\rm{r}}}{\left| {{g_i}} \right|^2}\left( {2{p_{\rm{r}}}{{\left| f \right|}^2} + \sigma _n^2} \right)\;$,  $\gamma  = 2{p_{\rm{s}}}{\left| {{h_i}} \right|^2}\left( {{p_{\rm{r}}}{{\left| f \right|}^2} + \sigma _n^2} \right) + p_{\rm{s}}^2{\left| {{h_i}} \right|^4} - 2\beta $.
\\ \vspace*{4pt}
\hrulefill
\end{figure*}

\section{Adaptive Rate Performance of Virtual Full Duplex Relaying}

 Designing adaptive VFD system can be done through the optimization of different system parameters. When proper  signals are adopted, $R_i$ can improve the rate of the second hop of the $i$th by increasing its transmit power. However, this will deteriorate the rate of the first hop of the $j$th path by increasing the interference level. Therefore, improper  signaling comes into the scene to reduce the interference. By increasing the asymmetry of the relay's transmit signal, i.e., by increasing the circularity coefficient, the relay can increase its power with reduced effect on the other relay. Moreover, optimizing the transmission phases duration can help to improve the system performance as we could use the phase which has the worst channel condition less frequently than the other phase. To maximize the achievable rate performance, we formulate the following optimization problem
\begin{align}
 &\mathop {\max }\limits_{\mathcal{C}_1,\mathcal{C}_2,\tau}\qquad  {\mathcal{R}_{{\rm{T}}}}\left( {\mathcal{C}_1,\mathcal{C}_2,\tau} \right)= \sum\limits_{i = 1}^2 {{\mathcal{R}_{{_i}}}\left( {{{\cal C}_1},{{\cal C}_2},\tau} \right)} \nonumber \\
&\;{\rm{s}}{\rm{.t}}{\rm{.}}\quad \quad \;\; 0\leq \mathcal{C}_1 \leq 1,\nonumber\\
& \qquad \qquad \; 0\leq \mathcal{C}_2 \leq 1,\nonumber\\
& \qquad \qquad \; 0\leq \tau \leq 1.  \label{convex_prob}
\end{align}
To evaluate the performance of the VFD system employing different improper  signals at each relay and operating with asymmetric time sharing, we solve the non-convex optimization problem in  \eqref{convex_prob} using grid search algorithm as finding a closed-form solution appears to be very challenging. Then, we  investigate the impact of different system parameters on the achievable rate performance of the VFD system through different numerical scenarios in the following subsections.
\subsection{Impact of Inter-Relay-Interference on the achievable Rate Performance}
\begin{figure}[!t]
\centering
\includegraphics[width=3.5in]{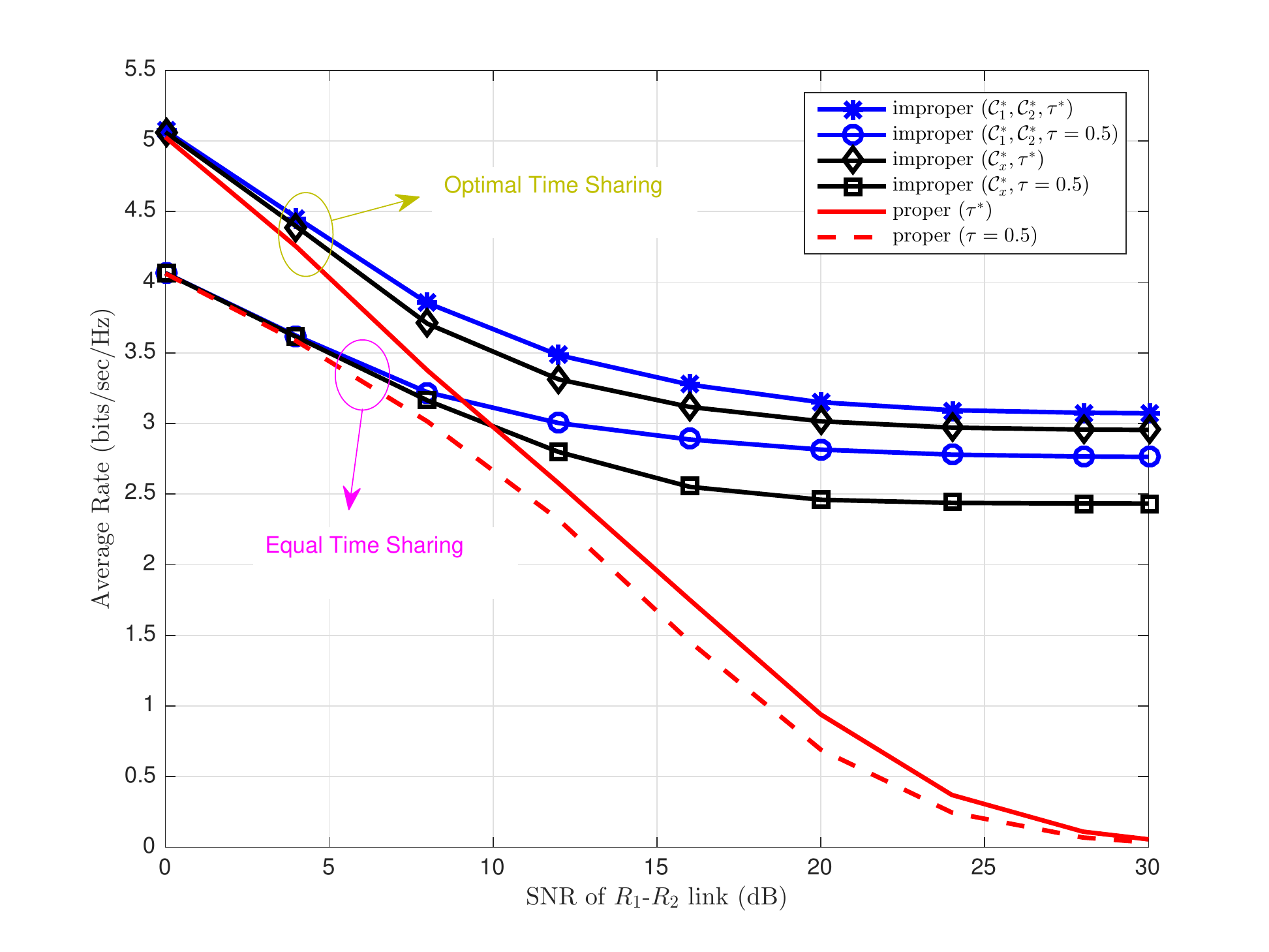}
\caption{The average achievable rates for different schemes versus the average IRI link gain $\sigma^2_f$ where $\sigma^2_{h_1}=5$ dB, $\sigma^2_{h_2}=20$ dB, $\sigma^2_{g_1}=15$ dB, $\sigma^2_{g_2}=10$ dB, $p_{\rm{s}}=p_{\rm{r}}=p_{\rm{max}}=1$ W, $\sigma^2_{n}=1$ and 10000 random channel realizations. }
\label{fig_2}
\end{figure}

\begin{figure}[!t]
\centering
\includegraphics[width=3.5in]{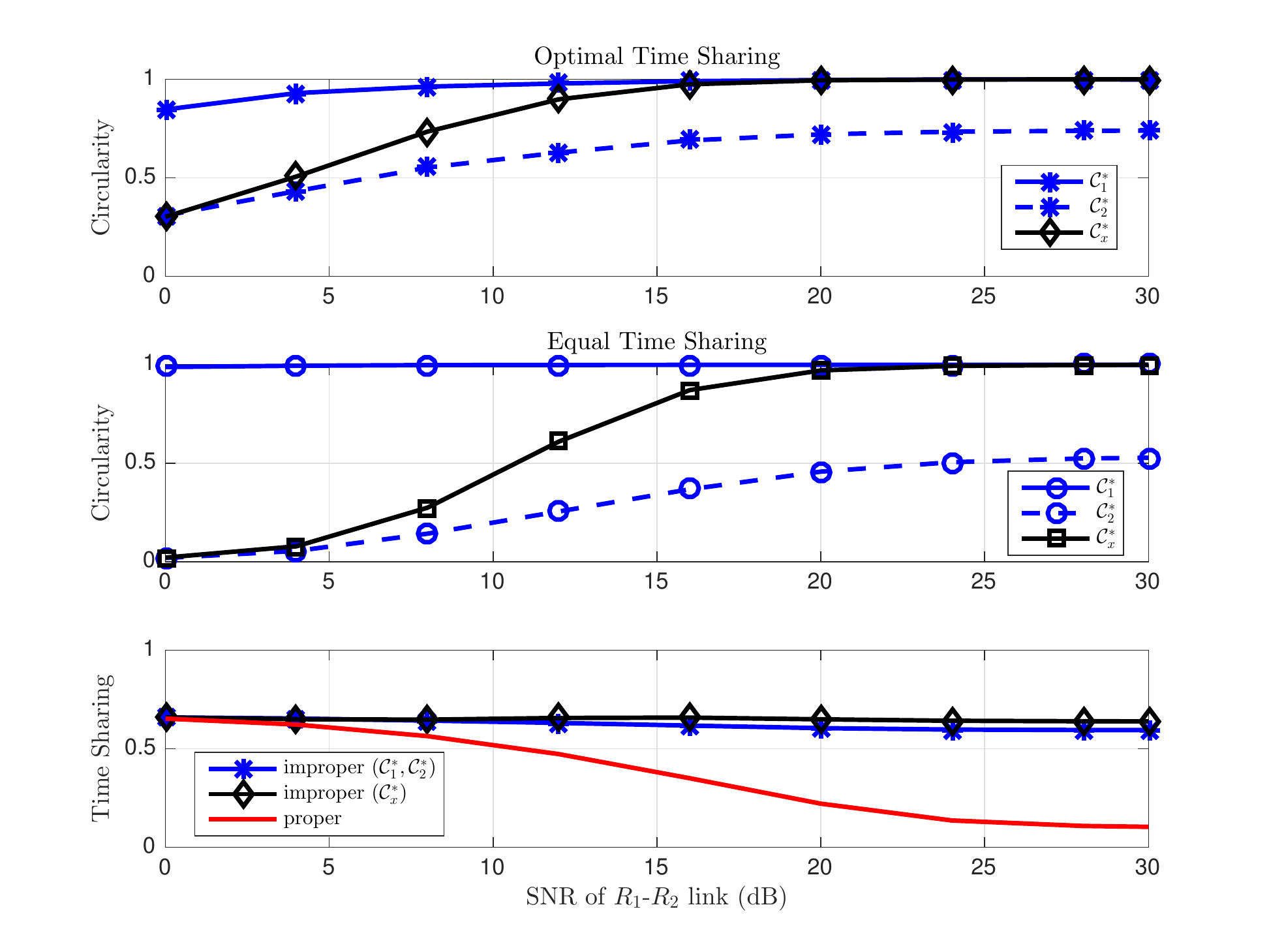}
\caption{The average optimal optimization parameters of the grid search for different schemes versus the average IRI link gain $\sigma^2_f$ where $\sigma^2_{h_1}=5$ dB, $\sigma^2_{h_2}=20$ dB, $\sigma^2_{g_1}=15$ dB, $\sigma^2_{g_2}=10$ dB, $p_{\rm{s}}=p_{\rm{r}}=p_{\rm{max}}=1$ W, $\sigma^2_{n}=1$ and 10000 random channel realizations. }
\label{fig_3}
\end{figure}

In this study, we aim to investigate the achievable rate performance of VFD with different designing scenarios versus the IRI. To this end, we consider the achievable rate performance of the following scenarios in Fig. \ref{fig_2}: different circularity coefficients at the relays and the same circularity, i.e., ${{\cal C}_i}={{\cal C}_j}=\mathcal{C}_{x}$, assuming equal and optimal time sharing strategies. Moreover, the optimized parameters  are shown in Fig. \ref{fig_3} for each scenario. 

First, we observe that the rate performance of the systems uses improper  signaling 
is superior to the proper ones in both time allocation strategies  at mid and high values of IRI. While, all scenarios tend to perform quite similar at low IRI. In this case, the IRI channel is weak  and each relay can transit with a higher power without affecting the other relay. As the IRI channel becomes stronger, the relay with proper  signaling cannot use higher power to avoid deteriorating the performance of the other relay. On the other hand, when  improper  signals are adopted, the relays can use higher powers and compensate their interference effect by increasing the signal asymmetry achieved by increasing the circularity coefficient as shown in Fig. \ref{fig_3}. As for adopting different circularity coefficients at both relays, the performance gain is notable compared with a single circularity coefficient as a result of asymmetric system setup, which is mostly the case in a general practical setup. Finally, at very high IRI, while proper performance severely deteriorate, improper signaling performance saturates at a constant rate value. 

As for the time sharing optimization impact on the achievable rate performance, has an apparent effect on improving the performance as it compensates the effect of bad links. To understand the time sharing solution performance, we consider separate hops cases as follows. Firstly, assume we deal with $S-R_1-D$ path only, we find the optimal time sharing solution reduces to be more than 0.5 in order to compensate the first link performance, which applies for all IRI range. Secondly, assume $S-R_2-D$ path only, the optimal time sharing solution reduces also to be more than 0.5 in order to compensate the first link performance for low IRI. On the other hand, the optimal time sharing solution reduces also to be less than 0.5 for high IRI to compensate the performance of $S-R_2$ hop. To have an idea about the time sharing solution trend of the VFD system with proper signaling, we consider the extreme cases of very low and high IRI values. At very low IRI values, both paths $S-R_1-D$ and $S-R_2-D$ needs more than 0.5 time sharing as seen in Fig.   \ref{fig_3}. On the other hand, at high IRI, the second path $S-R_2-D$ dominates the achievable rate performance, thus the time sharing solution reduces to its solution, which is less than 0.5 as shown in Fig.  \ref{fig_3}.  Furthermore, when adopting improper signaling, the effect of the IRI is significantly relieved via increasing the transmitted signal asymmetry which gives the chance for both good and bad transmission phases to be nearly used during the transmission of date from $S$ to $D$.



\subsection{Rate Performance of the VFD System under Maximally Improper Signaling}
\begin{figure}[!t]
\centering
\includegraphics[width=3.5in]{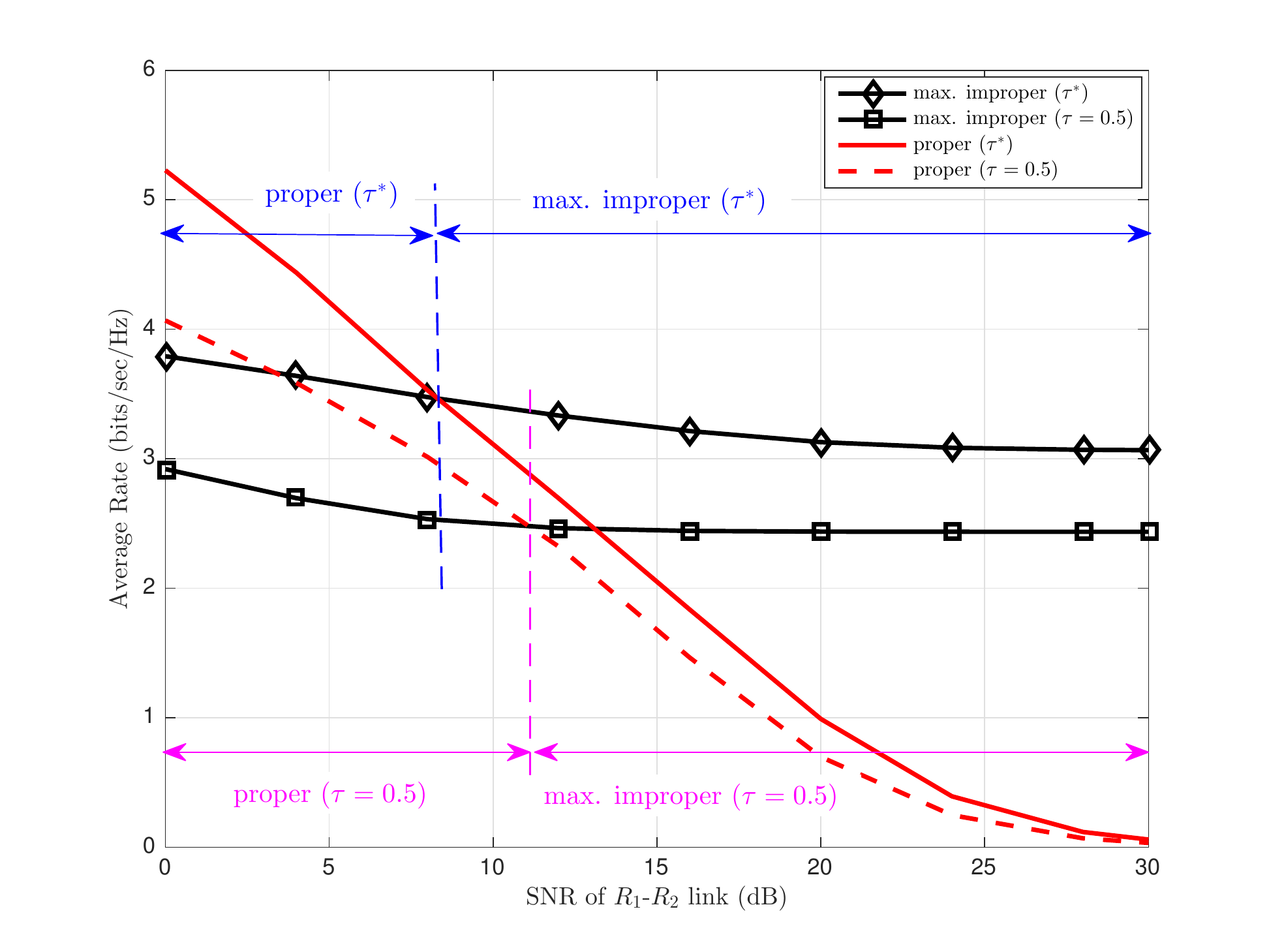}
\caption{The average achievable rates for different schemes under maximally improper signaling  versus the average IRI link gain $\sigma^2_f$ where $\sigma^2_{h_1}=5$ dB,  $\sigma^2_{h_2}=20$ dB, $\sigma^2_{g_1}=15$ dB, $\sigma^2_{g_2}=10$ dB, $p_{\rm{s}}=p_{\rm{r}}=p_{\rm{max}}=1$ W, $\sigma^2_{n}=1$ and 10000 random channel realizations. }
\label{fig_4}
\end{figure}

In Fig. \ref{fig_4}, we evaluate the rate performance of the VFD system when adopting only maximally improper signals at the relays, i.e., the signal lies either on the real or imaginary dimension. For each of the time sharing strategies, we find that the IRI range is divided into two regions according to the performance superiority. At low IRI, proper signaling is the best scheme, while the  maximally improper signal achieves the best performance at mid and high. These results matches the optimal circularity values in Fig. \ref{fig_3} where the circularity coefficients tend to be $1$ at high IRI. Moreover, using discrete design for the circularity coefficients reduces the complexity design and achieves a notable performance improvement. One can notice also from Fig. \ref{fig_3} that the boundary of the two regions is shifted to the right in case of equal time sharing which means that we could still use proper signaling for a higher value of the IRI when symmetric time sharing is considered and vice versa.

\subsection{Impact of Asymmetric Relays Location on the Achievable Rate Performance }
\begin{figure}[!t]
\centering
\includegraphics[width=3.5in]{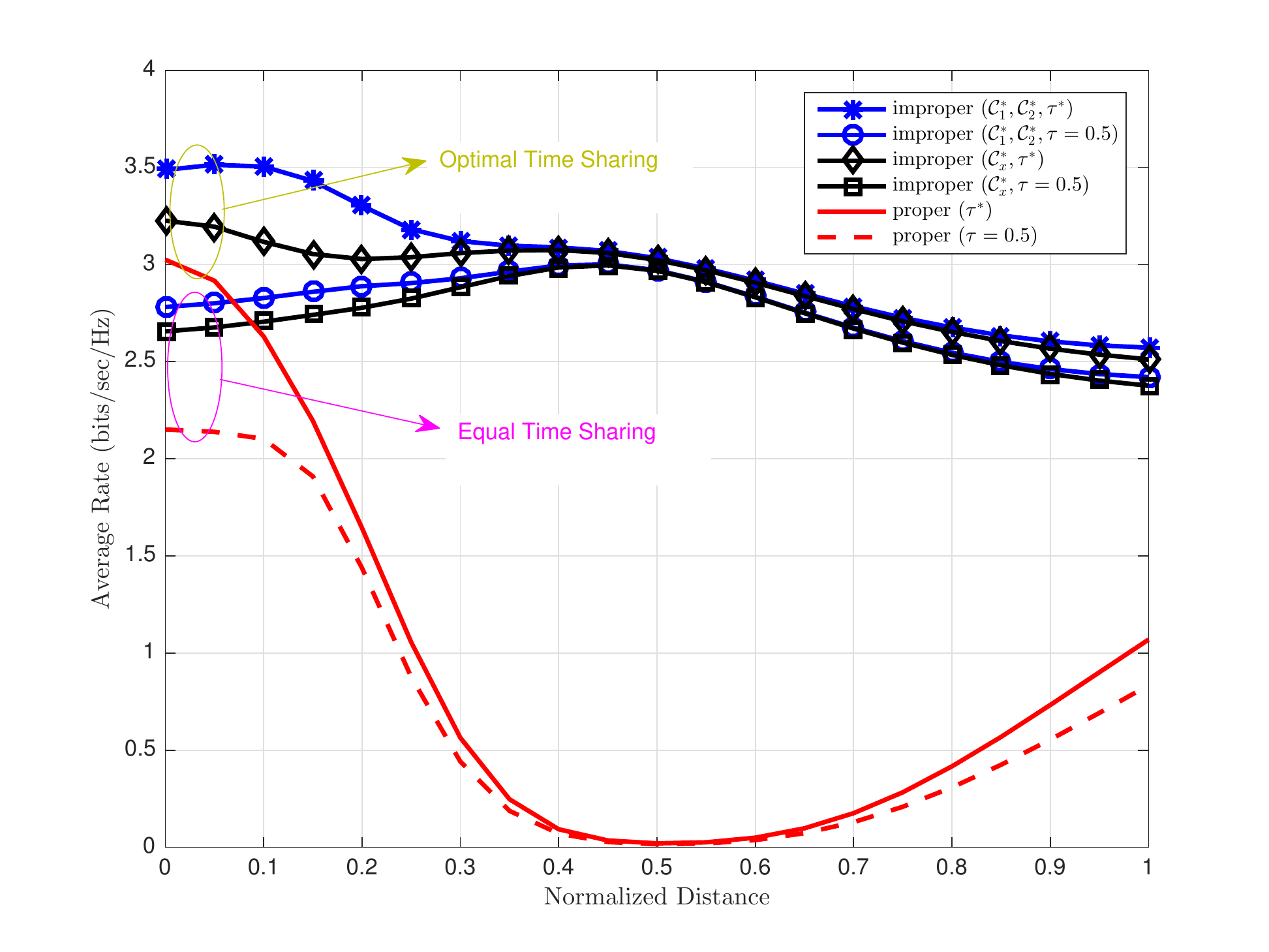}
\caption{The average achievable rates for different schemes versus the normalized distance between $S$ and $R_2$, $R_1$ lies in the middle distance between $S$ and $D$, where $p_{\rm{s}}=p_{\rm{r}}=p_{\rm{max}}=1$ W, $\sigma^2_{n}=1$, normalized vertical distance of $0.1$, shadowing
of 5 dB and a path loss exponent of 2  and 10000 random channel realizations. }
\label{fig_5}
\end{figure}

\begin{figure}[!t]
\centering
\includegraphics[width=3.5in]{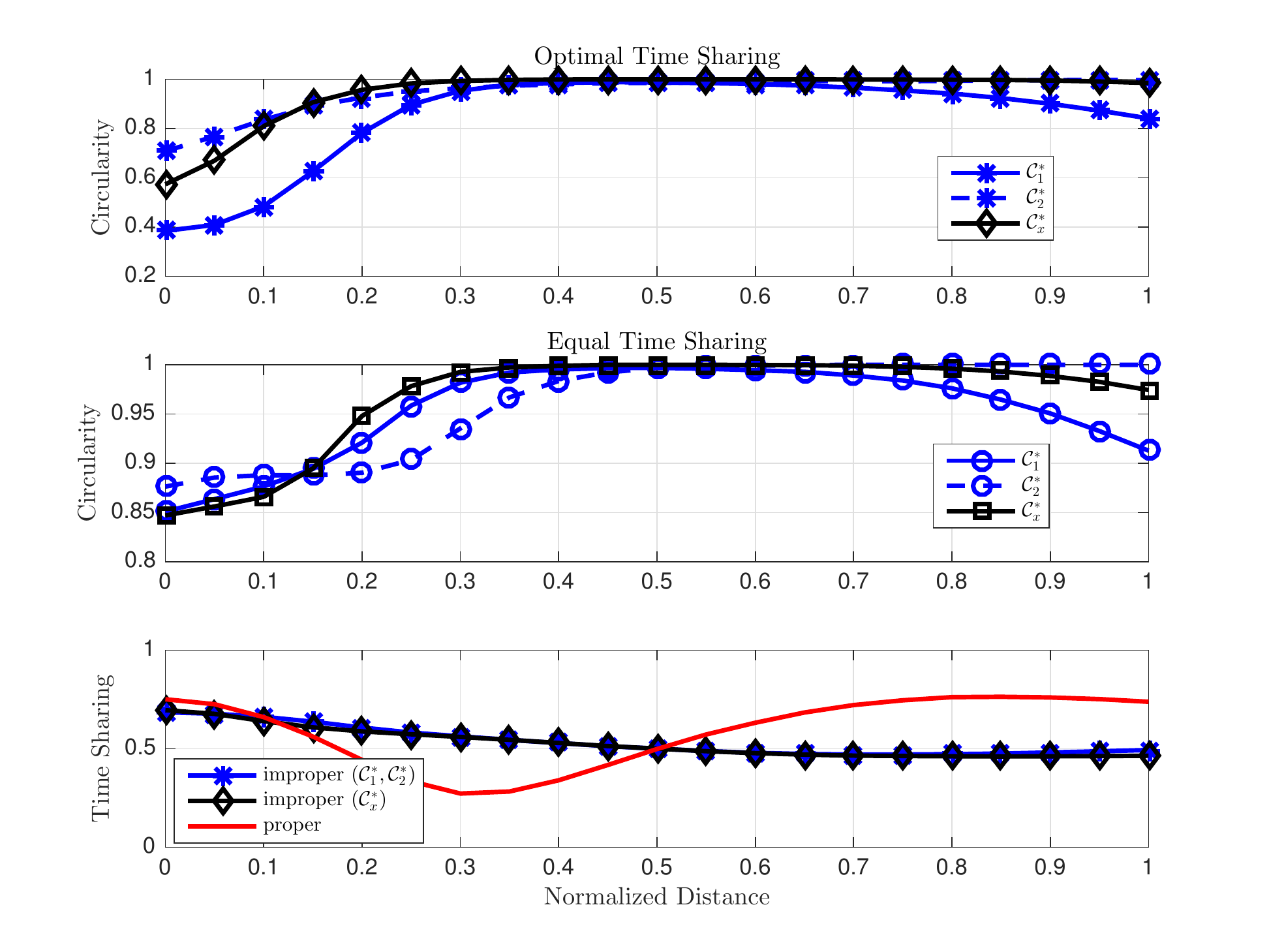}
\caption{The average optimal optimization parameters of the grid search for different schemes versus the normalized distance between $S$ and $R_2$, $R_1$ lies in the middle distance between $S$ and $D$,  where $p_{\rm{s}}=p_{\rm{r}}=p_{\rm{max}}=1$ W, $\sigma^2_{n}=1$, normalized vertical distance of $0.1$, shadowing
of 5 dB and a path loss exponent of 2 and 10000 random channel realizations. }
\label{fig_6}
\end{figure}

In this study, we investigate the effect of asymmetric relays location on the system performance with different optimized parameters. For this purpose, we plot the average achievable rate of different optimized systems in Fig. \ref{fig_5} versus the normalized distance between from $S$ to $R_2$\footnote{All distances in this simulation setup is normalized by the distance of the direct link between $S$ and $D$.}. $R_1$ is assumed to be fixed in the middle distance between $S$ and $D$. As for the average optimized design parameters, they are plotted in Fig. \ref{fig_6}. 

Firstly, the gain of employing improper signals with optimal time sharing over proper with/without optimal time sharing exists in all distance range. The maximum gain exists in the middle relays location, while the maximum rate is achieved when $R_2$ is located near to $S$. The benefit of using different circularity coefficient at both relays is observed whenever the VFD system becomes asymmetric in the structure, i.e., when $R_2$ is located near to the $S$ or $D$ which accordingly reflects on the time sharing gain. Also, the bigger time sharing gain is achieved when $R_2$ is located near to the source, which has low IRI and this confines with the results in Fig. \ref{fig_2}. For the specific distance value of $0.5$, all channel gains of the transmission links are symmetric and hence maximally improper with equal time sharing is optimal as it is clear from Fig. \ref{fig_6}. This is because there is no advantage of one transmission phase over the other one. This observation is also true here as the rates of all improper schemes nearly coincides at this point. The same observation is valid also for proper signaling which coincides to a very small rate value as this configuration yields the maximum IRI, which proper signaling fails to deal with even with optimal time sharing.

Also, from Fig. \ref{fig_6}, for proper signaling, as $R_2$ moves towards $R_1$, its direct channel with $S$ gets weaker and also the IRI from $R_1$ increases. Hence, the first hop of the second path deteriorates. As discussed previously, From the point of view of the $S-R_2-D$ path, $\tau$ should be decreased in order to maximize the effective rate of the second path by increasing the rate of its first hop which is used with $1-\tau$. However, while $R_2$ is moving to the right, the first hop of the $S-R_1-D$ path deteriorates and it tries to increase $\tau$ to boost the rate of the first hop as it is used with $\tau$. Furthermore, if $R_2$ is still in between $S$ and $R_1$, its first hop channel quality is still better than the first hop of the first path and hence, the overall optimization should favor the second path endeavor of decreasing $\tau$. This will be valid until certain distance (approx. $0.3$) in which the trade-off between the two paths can be obtained by solving the optimization problem according to the channel conditions of the simulation setup. As $R_2$ crosses the middle point, with the same aforementioned explanation, the optimization will favor the first path endeavor of increasing $\tau$ and hence the optimal time sharing parameter increases. For improper signaling, thanks to the power of adjusting the transmitted signal asymmetry, the effect of the IRI can be significantly relieved and the optimal time sharing parameter is around $0.5$, i.e., equal time sharing, which tries to make use of the two transmission phases similar to Fig. \ref{fig_3}.

%

\section{Conclusion}
In this paper, we consider a VFD relaying system under adopting improper signaling at the relays with different time sharing of the two transmission phases of the system. The achievable rates of the VFD system are developed based on the signal and time sharing asymmetries. Then, the VFD system parameters are optimized using a grid search algorithm in order to maximize the overall end-to-end achievable rate. Simulation results show the advantages of adopting improper signaling with optimized time sharing over proper with/without equal time sharing, at mid and high inter-relay interference (IRI) values. Moreover, the benefits of employing two different circularity coefficients at the relays can be reaped if the VFD system setup is asymmetric.

Future research lines can consider a more sophisticated and rigorous theoretical analysis of the optimization problem in order to design the circularity and time sharing coefficients. Also, the transmit power of the relays can be tuned to improve the overall rate performance of the VFD system.   
\appendices

\bibliographystyle{IEEEtran}

\bibliography{IEEEabrv,gaafar_ref_ICCW'17}

\begin{thebibliography}{10}
\providecommand{\url}[1]{#1}
\csname url@samestyle\endcsname
\providecommand{\newblock}{\relax}
\providecommand{\bibinfo}[2]{#2}
\providecommand{\BIBentrySTDinterwordspacing}{\spaceskip=0pt\relax}
\providecommand{\BIBentryALTinterwordstretchfactor}{4}
\providecommand{\BIBentryALTinterwordspacing}{\spaceskip=\fontdimen2\font plus
\BIBentryALTinterwordstretchfactor\fontdimen3\font minus
  \fontdimen4\font\relax}
\providecommand{\BIBforeignlanguage}[2]{{%
\expandafter\ifx\csname l@#1\endcsname\relax
\typeout{** WARNING: IEEEtran.bst: No hyphenation pattern has been}%
\typeout{** loaded for the language `#1'. Using the pattern for}%
\typeout{** the default language instead.}%
\else
\language=\csname l@#1\endcsname
\fi
#2}}
\providecommand{\BIBdecl}{\relax}
\BIBdecl

\bibitem{Yang_HXM_COMMAG09}
Y.~Yang, H.~Hu, J.~Xu, and G.~Mao, ``Relay technologies for {WiMax} and
  {LTE}-advanced mobile systems,'' \emph{IEEE Commun. Mag.}, vol.~47, no.~10,
  pp. 100--105, Oct. 2009.

\bibitem{amin2012adaptive}
O.~Amin and M.~Uysal, ``Adaptive power loading for multi-relay {OFDM}
  regenerative networks with relay selection,'' \emph{{IEEE} Trans. Commun.},
  vol.~60, no.~3, pp. 614--619, Mar. 2012.

\bibitem{amin2011optimal}
------, ``Optimal bit and power loading for amplify-and-forward cooperative
  {OFDM} systems,'' \emph{{IEEE} Trans. Wireless Commun.}, vol.~10, no.~3, pp.
  772--781, Mar. 2011.

\bibitem{sabharwal2014band}
A.~Sabharwal, P.~Schniter, D.~Guo, D.~W. Bliss, S.~Rangarajan, and R.~Wichman,
  ``In-band full-duplex wireless: Challenges and opportunities,'' \emph{{IEEE}
  J. Sel. Areas Commun.}, vol.~32, no.~9, pp. 1637--1652, Sep. 2014.

\bibitem{Rankov_W_JSAC07}
B.~Rankov and A.~Wittneben, ``Spectral efficient protocols for half-duplex
  fading relay channels,'' \emph{IEEE J. Select. Areas Commun.}, vol.~25,
  no.~2, pp. 379--389, Feb. 2007.

\bibitem{ju2009catching}
H.~Ju, E.~Oh, and D.~Hong, ``Catching resource-devouring worms in
  next-generation wireless relay systems: two-way relay and full-duplex
  relay,'' \emph{{IEEE} Commun. Mag.}, vol.~47, no.~9, pp. 58--65, Sep. 2009.

\bibitem{gaafar2017improper}
\BIBentryALTinterwordspacing
M.~Gaafar, O.~Amin, R.~F. Schaefer, and M.-S. Alouini, ``{Improper Signaling in
  Two-Path Relay Channels},'' in \emph{Proc. IEEE Int. Conf. Communications
  (ICC) Workshops. Submitted}, Paris, France, May. 2017. [Online]. Available:
  \url{https://arxiv.org/abs/1612.05743}
\BIBentrySTDinterwordspacing

\bibitem{Lameiro_SS_WCL15}
C.~Lameiro, I.~Santamaria, and P.~Schreier, ``Benefits of improper signaling
  for underlay cognitive radio,'' \emph{IEEE Wireless Commun. Lett.}, vol.~4,
  no.~1, pp. 22--25, Feb. 2015.

\bibitem{amin2017overlay}
O.~Amin, W.~Abediseid, and M.-S. Alouini, ``Overlay spectrum sharing using
  improper gaussian signaling,'' \emph{{IEEE} J. Sel. Areas Commun.}, vol.~35,
  no.~1, pp. 50--62, Jan. 2017.

\bibitem{gaafar2017underlay}
M.~Gaafar, O.~Amin, W.~Abediseid, and M.-S. Alouini, ``Underlay spectrum
  sharing techniques with in-band full-duplex systems using improper {G}aussian
  signaling,'' \emph{{IEEE} Trans. Wireless Commun.}, vol.~16, no.~1, pp.
  235--249, Jan. 2017.

\bibitem{kurniawan2015improper}
E.~Kurniawan and S.~Sun, ``Improper gaussian signaling scheme for the
  {Z}-interference channel,'' \emph{{IEEE} Trans. Wireless Commun.}, vol.~14,
  no.~7, pp. 3912--3923, Jul. 2010.

\bibitem{gaafar2016letter}
M.~Gaafar, O.~Amin, A.~Ikhlef, A.~Chaaban, and M.~S. Alouini, ``On alternate
  relaying with improper gaussian signaling,'' \emph{{IEEE} Commun. Lett.},
  vol.~20, no.~8, pp. 1683--1686, Aug 2016.

\bibitem{neeser1993proper}
F.~D. Neeser and J.~L. Massey, ``Proper complex random processes with
  applications to information theory,'' \emph{{IEEE} Trans. Inf. Theory},
  vol.~39, no.~4, pp. 1293--1302, Jul. 1993.

\bibitem{zeng2013transmit}
Y.~Zeng, C.~M. Yetis, E.~Gunawan, Y.~L. Guan, and R.~Zhang, ``{Transmit
  optimization with improper Gaussian signaling for interference channels},''
  \emph{{IEEE} Trans. Signal Process.}, vol.~61, no.~11, pp. 2899--2913, Jun.
  2013.

\end{thebibliography}

\vfill

\end{document}